\documentclass[aps,prl,twocolumn,reprint, showpacs]{revtex4-1}
\pdfoutput=1
\usepackage[pdftex]{graphicx}
\usepackage{natbib}
\usepackage[cmex10]{amsmath}
\usepackage{color}
\usepackage{soul}

\usepackage[]{caption}
\usepackage[font=footnotesize]{subfig}

\begin{document}
\title{Active Cloaking}

\author{Michael~Selvanayagam and George.~V.~Eleftheriades }
\affiliation{The Edward S. Rogers Department
of Electrical and Computer Engineering, University of Toronto, Toronto}

\date{\today}

\begin{abstract}
Electromagnetic cloaking refers to the ability to prevent an object from scattering an incident electromagnetic field. This has been accomplished in recent works by routing the incident field around the object or by changing the scattering properties of the object itself through specially designed materials, surfaces or guiding structures. In this letter, we introduce a new way of cancelling the electromagnetic scattering of an object by using an array of sources.  We show that by superimposing magnetic and electric surface current densities at the boundary of an object, the scattered fields from that object can be cancelled.  These magnetic and electric surface currents can be  discretized into electric and magnetic dipoles which are physically implementable by straight and loop wire antennas. Finally we confirm our results using numerical simulations for cloaking a dielectric and a metallic cylinder by means of a thin array of such wire antennas.
\end{abstract}


\maketitle

The concept of electromagnetic cloaking marks recent important advances in the ability to control electromagnetic waves. Over the past five years many different approaches to cloaking have been developed, demonstrating the variety of ways to hide an object from an incident electromagnetic wave. One of the first approaches was based on transformation optics \cite{schurig_etal_2006}. To create a cloak using transformation optics, an incident electromagnetic wave can be bent around an object using an anisotropic and inhomogeneous material. This approach originates from the equivalence between geometrical transformations and material parameters in Maxwell's equations. Transformation-optics cloaking is completely general and can hide an arbitrary object of any size. The challenge however resides in the ability to physically realize the anisotropic and inhomogeneous material using metamaterial techniques \cite{schurig_etal_2006}.  Moreover, depending on the size of the object, the corresponding cloak can be substantially thick.

Other approaches to the cloaking problem consider the specific geometry/composition of the object being hidden. One example is plasmonic cloaking which demonstrates the ability to cancel the dipolar scattering of an electromagnetically small object.  This is achieved by covering the object with a plasmonic material or surface.  The field incident on the object and the plasmonic surface generates two oppositely phased dipole moments thus cancelling the dipolar fields scattered of the object itself \cite{Alu_Engheta_2008}. Here the material is designed specifically for the composition of the object being hidden.  Moreover, it should be noted that higher order scattering terms (quadrupolar, etc.) are not canceled with this method.

Yet another approach to cloaking allows for the use of a network of transmission-lines or waveguides to route the incident field through or around the object being hidden \cite{Alitalo_Tretyakov_2011}.  While this approach can cloak an arbitrarily sized object, it is also dependent on the object being hidden and requires a thick waveguide surrounding the structure.

Finally, specially designed anisotropic surfaces can be used to minimize scattering in specific directions (such as the forward scattering) by guiding the wave around the object using the anisotropic surface itself \cite{Kildal_etal_1996}.

As described above, all of these cloaks depend on the use of materials/waveguides to bend, guide or scatter the incident electromagnetic wave in such a way as to remove the scattered field. The common thread uniting all of these cloaks is that they are passive. By passive, we mean cloaks which simply alter the existing incident or scattered field without introducing any other fields into the problem.  In this letter, we approach the problem of cloaking differently; instead of using a specifically designed material or waveguide to hide an object, we demonstrate cloaking by means of an array of sources. These sources are specifically designed to introduce a set of fields to cancel out the scattered field of the object being cloaked. 

\begin{figure*}[!t]
\centering
\subfloat[]
{
	\includegraphics[clip=true, trim= 4cm 4.5cm 7cm 3cm, scale=0.25]{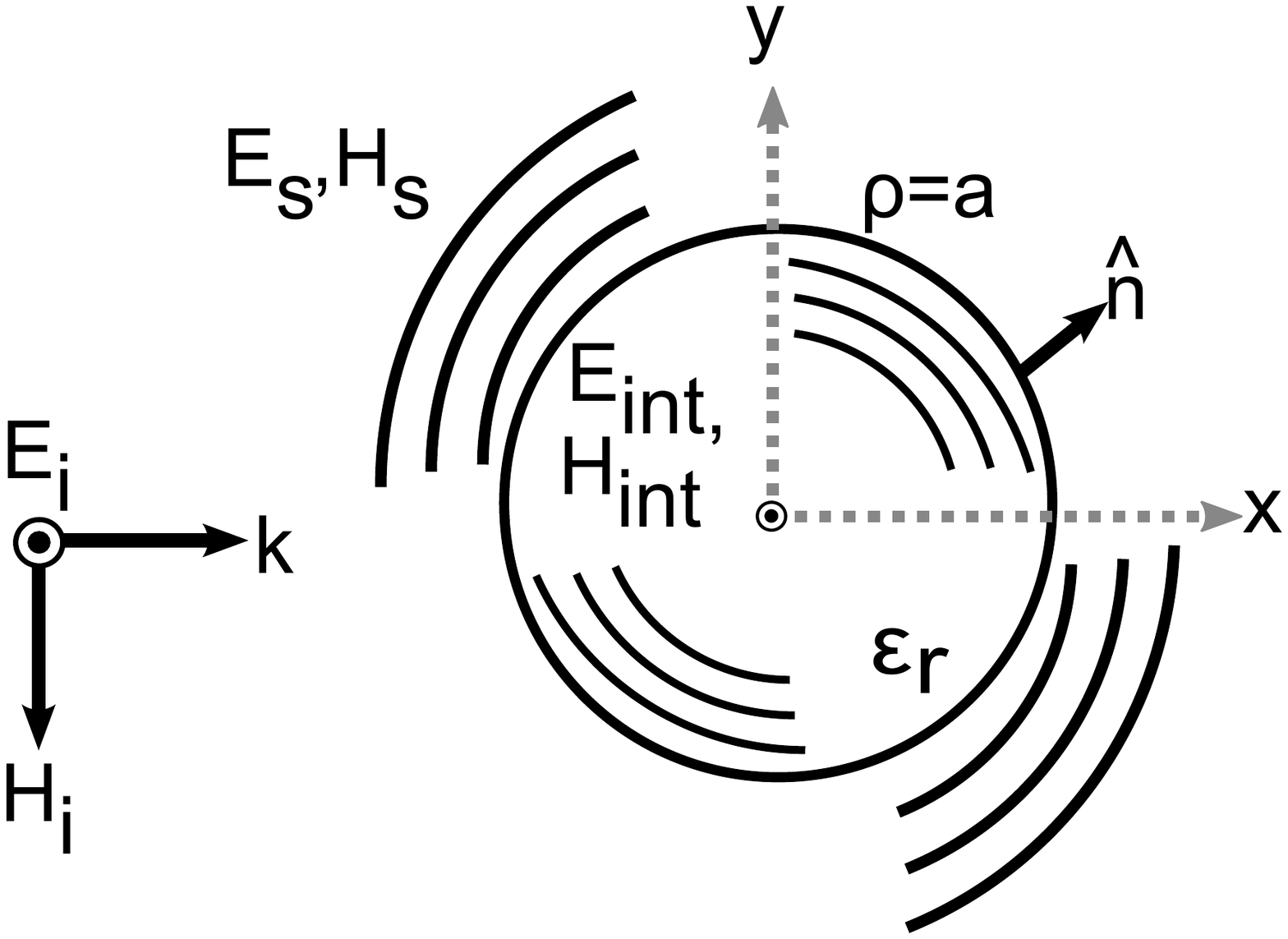}
	\label{fig:cylScat}
}
\subfloat[]
{
	\includegraphics[clip=true, trim= 5cm 4.5cm 3cm 3cm, scale=0.25]{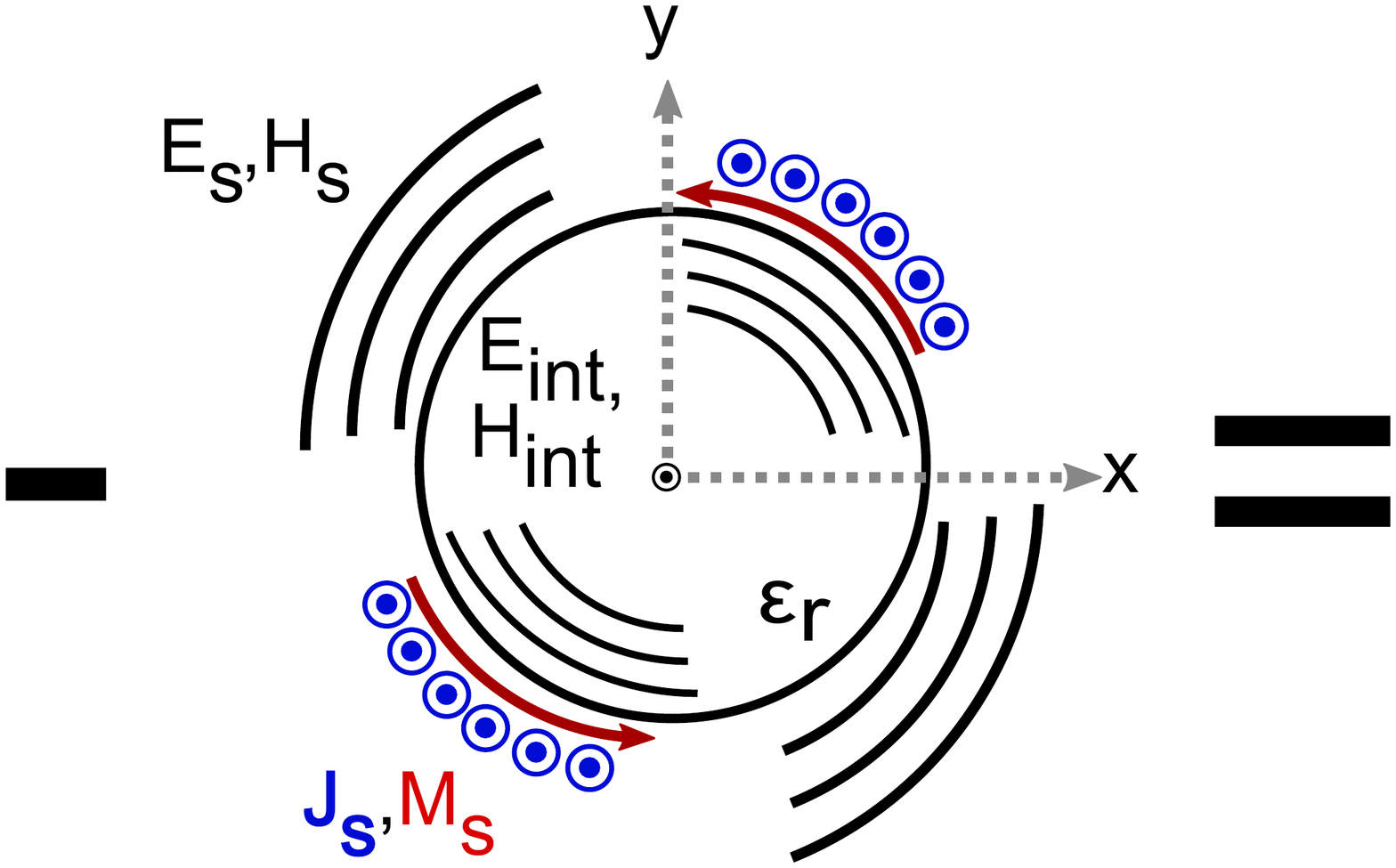}
	\label{fig:JsMs}
}
\subfloat[]
{
	\includegraphics[clip=true, trim= 4cm 4.5cm 7cm 3cm, scale=0.25]{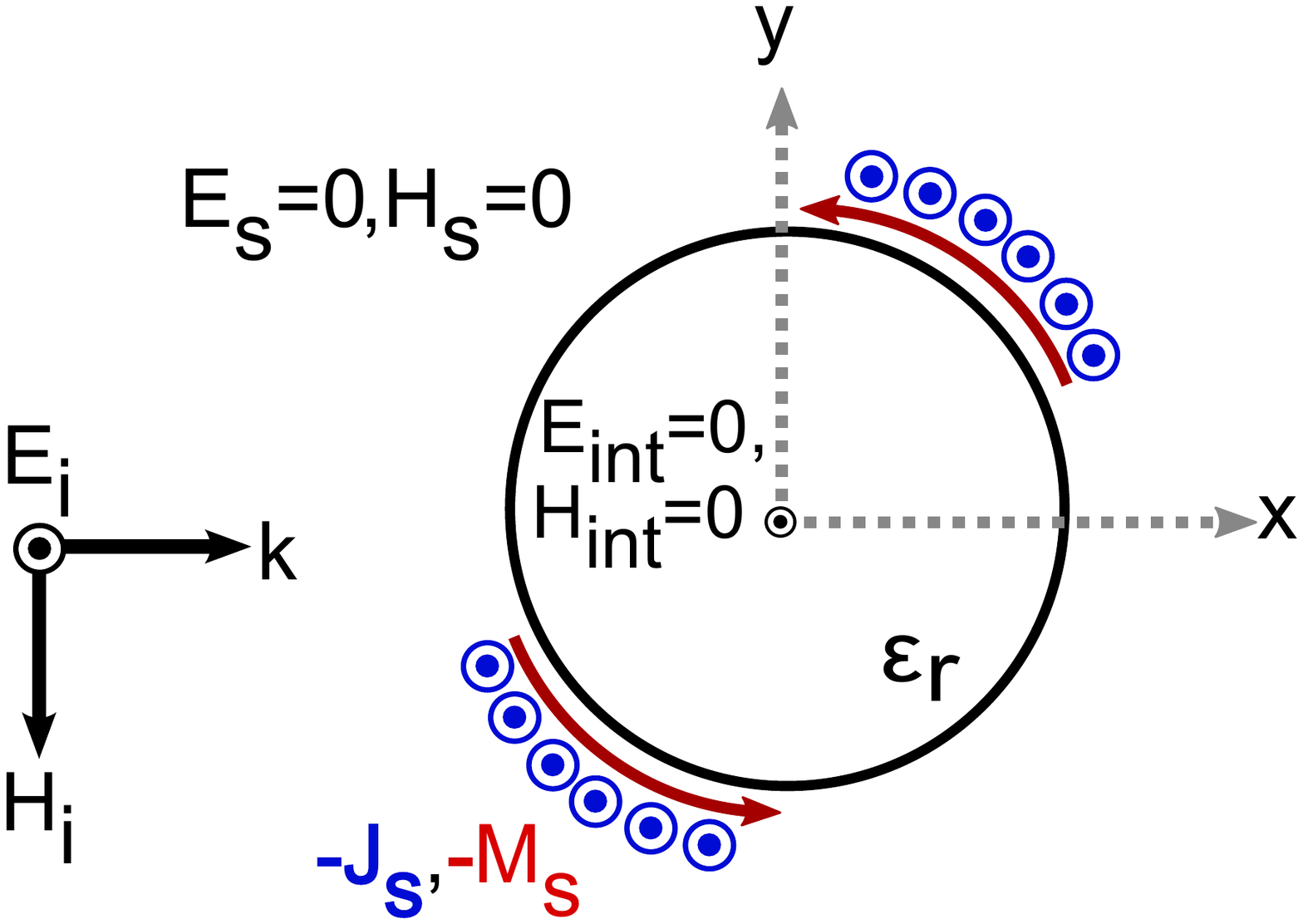}
	\label{fig:cloakSchem}
}
\caption{ \protect\subref{fig:cylScat} Scattering of a plane wave off of a cylindrical object  \protect\subref{fig:JsMs} An equivalent scenario where the incident plane wave is replaced with electric and magnetic current densities on the surface of the cylindrical object \protect\subref{fig:cloakSchem} The superposition of \protect\subref{fig:cylScat} and \protect\subref{fig:JsMs} such that the scattered and interior fields are cancelled out.}
\label{fig:ActiveCloakIdea}
\end{figure*}

The geometry of our problem is shown in Fig.~\ref{fig:cylScat} and consists of a cylindrical object located at the origin with radius $\rho=a$ and  with a dielectric constant given by $\epsilon_r$ (while we work with dielectrics, this method can be extended to other materials such as magnetic materials).  For simplicity we will work in a cylindrical geometry assuming infinite extent along the $z$-axis, though all of the concepts presented herein can be extended to fully three-dimensional geometries.  

As also shown in Fig.~\ref{fig:cylScat}  we have an incident directed plane wave travelling along the $x$-axis denoted by its electric field component $\mathbf{E_i}=e^{-jkx}\mathbf{\hat{z}}$, where $k$ is the free space wavenumber. Without any loss of generality, we will assume that this incident field is polarized along the $z$-axis (TE-polarization).  This incident plane wave is scattered by the cylindrical object creating a scattered electric field outside the object, $\mathbf{E_s}$ and an interior field, $\mathbf{E_{int}}$.  A similar situation exists for the magnetic field with the magnetic field separated into incident, scattered and interior fields ($\mathbf{H_i}=(-1/\eta) e^{-jkx}\mathbf{\hat{y}}$, $\mathbf{H_s}$, $\mathbf{H_{int}}$, where $\eta$ is the free space wave impedance).

This situation can be described as a boundary value problem determined by the boundary condition at the interface between free-space and the cylindrical object. 
At the boundary, $\rho=a$, the boundary conditions at the interface enforce the continuity of the tangential components of the electric and magnetic fields. These boundary conditions can be expressed as,
\begin{equation}
\label{eq:EBC}
\mathbf{\hat{n}} \times \mathbf{E_i} + \mathbf{\hat{n}} \times \mathbf{E_s} = \mathbf{\hat{n}} \times \mathbf{E_{int}}, 
\end{equation}
\begin{equation}
\label{eq:HBC}
\mathbf{\hat{n}} \times \mathbf{H_i} + \mathbf{\hat{n}} \times \mathbf{H_s}= \mathbf{\hat{n}} \times \mathbf{H_{int}}, 
\end{equation}
where $\mathbf{\hat{n}}$ is the outward unit vector to the surface of the object.  

Let us now imagine a different scenario illustrated in Fig.~\ref{fig:JsMs} which consists of the same cylindrical object as before.  We now consider what happens when the incident field is removed while attempting to maintain the same scattered and interior fields determined before. To do this we inspect the boundary conditions given in \eqref{eq:EBC} and \eqref{eq:HBC}.  With the incident fields removed, we can notice that the boundary condition at $\rho=a$ is no longer satisfied as the fields are now discontinuous.  However this can be remedied by noting that the general boundary conditions for the tangential electric and magnetic fields are satisfied by the presence of sources impressed at the boundary in the case of discontinuity.  For the tangential electric field this gives rise to a magnetic surface current density equal to the discontinuity of the tangential electric field at the boundary. Likewise, for the tangential magnetic field a surface electric current density equal to the discontinuity of the tangential magnetic field is impressed at the boundary \cite{Harrington}. This yields a set of boundary conditions for the tangential electric and magnetic fields at $\rho=a$ as,
\begin{equation}
\label{eq:MDef}
\mathbf{M_s}= - \mathbf{\hat{n}}  \times \left[\mathbf{ E_{int} - E_s }\right],
\end{equation}
\begin{equation}
\label{eq:JDef}
\mathbf{J_s}= \mathbf{\hat{n}}  \times \left[\mathbf{H_{int} - H_s} \right],
\end{equation}
The value of these surface current densities is dictated by \eqref{eq:MDef} and \eqref{eq:JDef} where we can see that both the magnetic and electric surface current densities are equal to the difference between the scattered and interior fields.  From \eqref{eq:EBC} and \eqref{eq:HBC} we can observe that the difference between the scattered and interior fields is simply the incident field. Therefore, this determines the magnetic and electric surface current densities as,
\begin{equation}
\label{eq:M}
\mathbf{M_s}= - \mathbf{\hat{n}}  \times \mathbf{E_i},
\end{equation}
\begin{equation}
\label{eq:J}
\mathbf{J_s}= \mathbf{\hat{n}}  \times \mathbf{H_i},
\end{equation}
which are related to the incident field which was defined as a plane wave in our example. We can now interpret these surface current densities as a set of equivalent sources which generate the scattered and interior fields of the cylindrical object in lieu of the incident field \cite{Harrington}.  We note that the scattered and interior fields are generated by the surface current densities, \eqref{eq:M} and \eqref{eq:J}  in the presence of the object.
With these two scenarios in mind we can look into cancelling the scattered fields outside of the cylindrical object  by superimposing the scenarios described in Fig.~\ref{fig:cylScat} with Fig.~\ref{fig:JsMs} as shown in Fig.~\ref{fig:cloakSchem}.  In this situation we have an incident field, $\mathbf{E_i,H_i}$, impinging upon the cylindrical object which generates the scattered and interior fields. However, we also have a set of magnetic and electric surface current densities surrounding the object, $\mathbf{M_{cloak}}$ and $\mathbf{J_{cloak}}$ respectively, with the magnetic and electric surface current densities given by the negatives of \eqref{eq:M} and \eqref{eq:J} with  $\mathbf{M_{cloak}=-M_s}$ and $\mathbf{J_{cloak}=-J_s}$.
Here $\mathbf{M_{cloak}}$ and $\mathbf{J_{cloak}}$ are $180^{\circ}$ out of phase with the incident field. Because of this phasing, the total field outside the object is simply the incident field by superposition.  This is because the scattered fields generated by the incident field are cancelled by the scattered fields generated by the magnetic and electric surface current densities, $\mathbf{M_{cloak}}$ and $\mathbf{J_{cloak}}$. Likewise, the interior field vanishes.  These electric and magnetic surface currents are used  to restore the original incident field pattern. With these sources in place around the object, the scenario in Fig.~\ref{fig:cloakSchem} is equivalent to the object being cloaked as all that is left after superimposing the two cases in Fig.~\ref{fig:cylScat} and Fig.~\ref{fig:JsMs} are the incident fields $\mathbf{E_i,H_i}$.  Thus by covering an object with a set of appropriately phased electric and magnetic surface current densities we can also arrive at the cloaking phenomenon demonstrated by other methods. Note that we do not rely on the use of materials or waveguides to route the fields around the object, we instead use equivalent sources to actively cancel the scattered fields, hence the designation of our cloak as ``active". 

In the rest of this letter we examine how a set of sources that can reproduce an electric or a magnetic surface current density can be realized.

While the surface current densities described by $\mathbf{M_{cloak}}$ and $\mathbf{J_{cloak}}$ are continuous functions on the boundary of the object being hidden, a simple way of approximating such a distribution is by using electric and magnetic dipoles.  These electric and magnetic dipoles can be arranged around the boundary of the object in such a way as to implement a discrete version of a magnetic or an electric surface current. This then raises two questions, 1) How many electric or magnetic dipoles are needed? and 2) What are the weights required on each dipole? We begin by answering the first question. 

\begin{figure}[!t]
\centering
\subfloat[]
{
	\includegraphics[clip=true, trim= 0cm 0cm 0cm 0cm, scale=0.2]{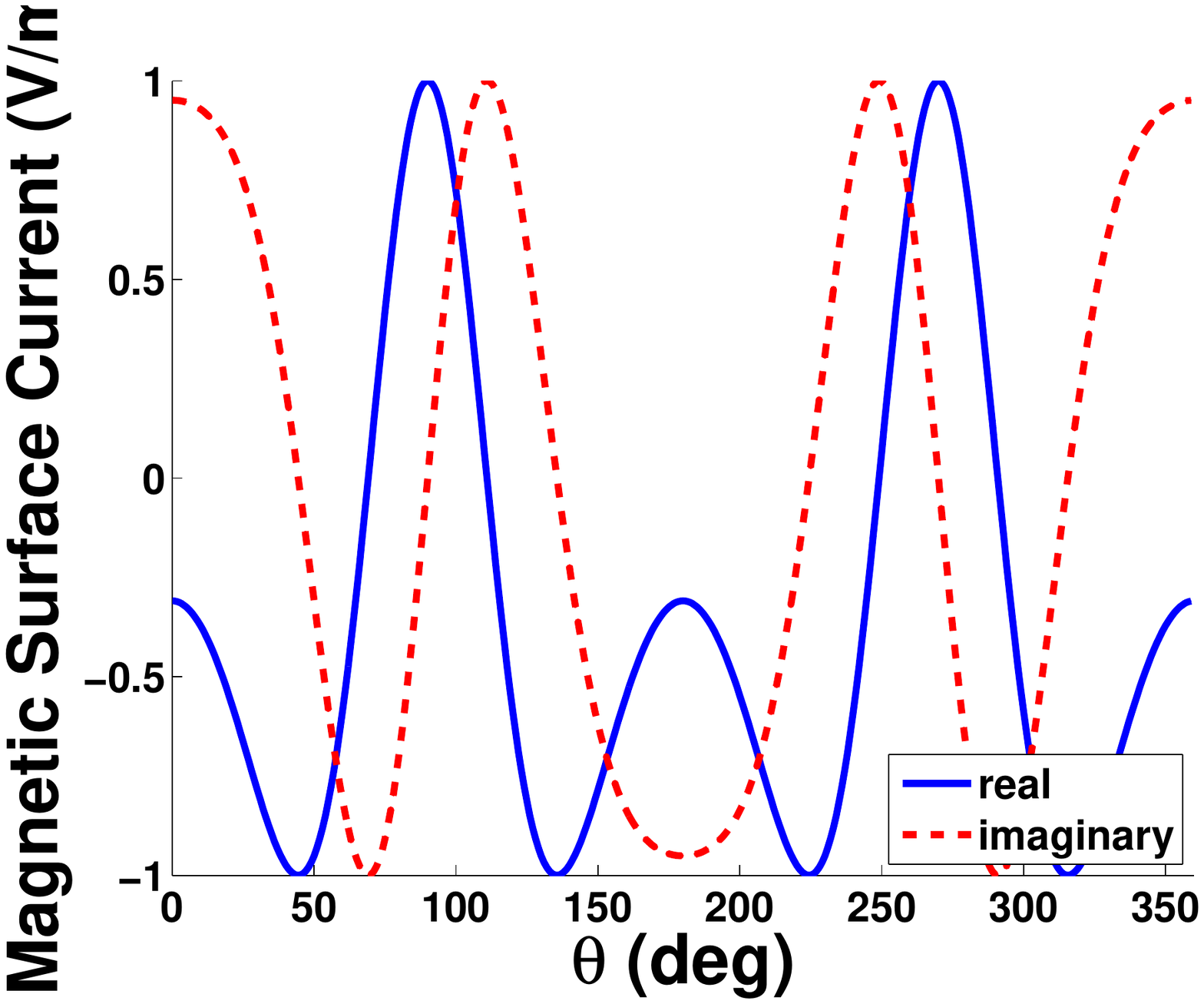}
	\label{fig:Ms}
}
\subfloat[]
{
	\includegraphics[clip=true, trim= 0cm 0cm 0cm 0cm, scale=0.2]{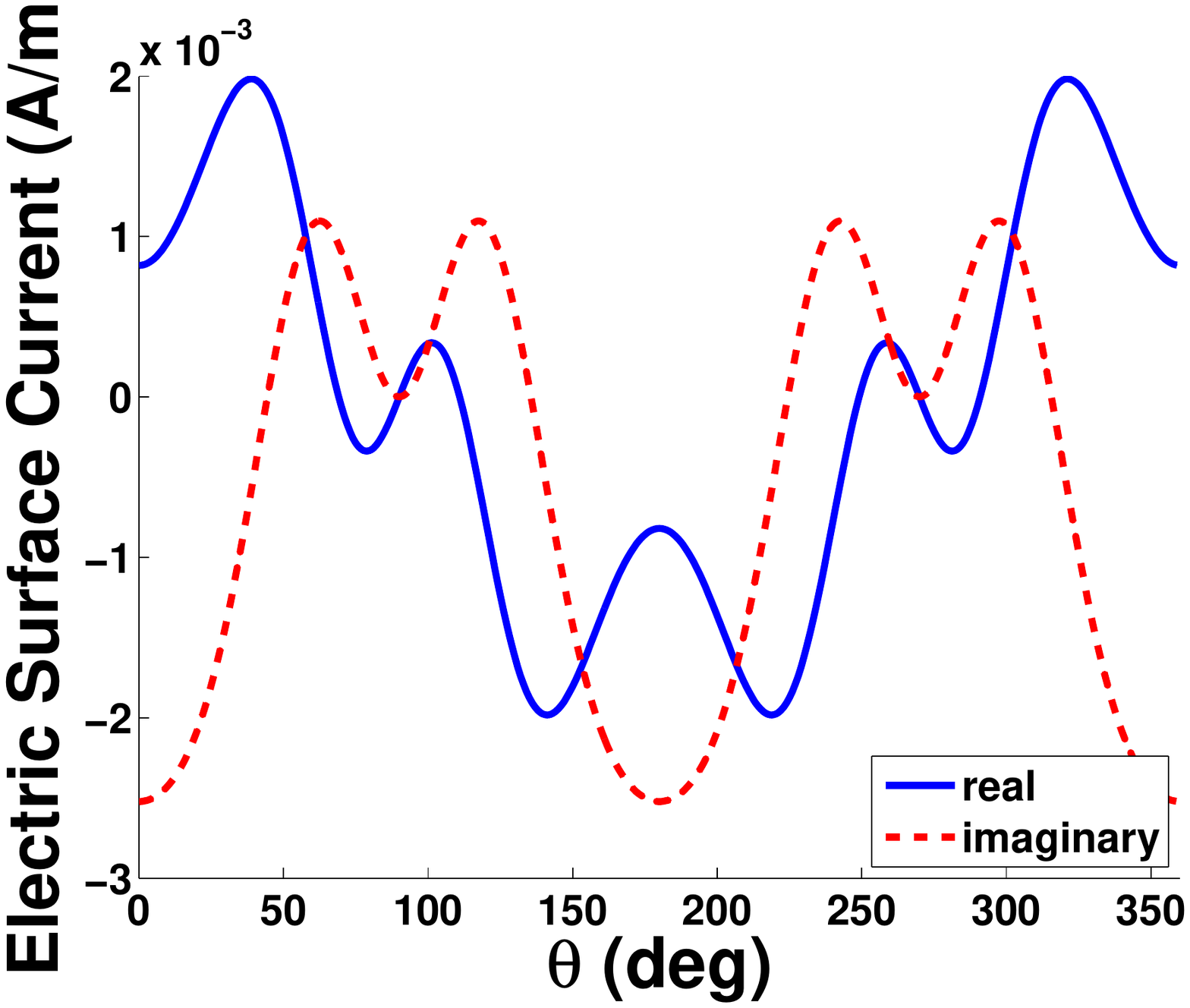}
	\label{fig:Js}
}
\caption{ \protect\subref{fig:Ms} The equivalent magnetic surface current on the boundary of the cylinder for a TE-polarized plane wave. \protect\subref{fig:Js} The equivalent electric surface current on the boundary of the cylinder for a TE-polarized plane wave.}
\label{fig:IncField}
\end{figure}

Keeping in mind that we are working in a cylindrical geometry, the boundary of our object is circular.  For an incident plane wave travelling in the $\hat{x}$ direction with a $\hat{z}$ directed electric field, the magnetic and electric currents are,
\begin{equation}
\label{eq:Mplane}
\mathbf{M_{cloak}}=-e^{-j k a\cos\theta} \mathbf{\hat{\theta}}
\end{equation}
\begin{equation}
\label{eq:Jplane}
\mathbf{J_{cloak}}=\frac{\cos\theta}{\eta} \left[e^{-j k a\cos\theta}\right] \mathbf{\hat{z}}
\end{equation}
and are plotted in Fig.~\ref{fig:IncField} for a cylinder with radius $a=0.7~\lambda$.  On a circular boundary these current distributions are periodic and can be decomposed into a set of discrete sinusoidal functions, $e^{j m \theta }$ where $m$ is the order of the sinusoid. By finding the highest non-zero sinusoid needed to reconstruct this surface current distribution, $m=M$ we can then sample the surface current distribution with $N=2M$ sources placed at equal angular spacings around the circular boundary as per the Nyquist theorem \cite{Oppenheim}.  This is done by taking \eqref{eq:Mplane} and \eqref{eq:Jplane} and applying a discrete Fourier series (FFT) to determine $M$.  For a cylinder with radius $a=0.7\lambda$ we find that we need $N=20$ samples to reproduce the magnetic surface current distribution given in \eqref{eq:Mplane} and $N=24$ samples to reproduce the electric surface current distribution given in \eqref{eq:Jplane}. Thus we need 20 magnetic dipoles and 24 electric dipoles each, surrounding the object. We note that this method can be carried out for other geometric boundaries and other incident field patterns. 
With the number of electric and magnetic dipoles known, the weights of each dipole can be determined by converting the magnetic and electric surface current densities into electric and magnetic dipole moments. These dipole moments are directed along the same direction as their corresponding surface current densities. For the magnetic dipole moment its magnitude $\mathbf{p_m}$ can be related to the magnetic surface current density by,
\begin{equation}
\label{eq:MWeight}
|\mathbf{p_m}|=|\mathbf{M_s}| h l,
\end{equation}
where $h$ is the height of the cylindrical volume that $\mathbf{M_s}$ resides on and $l$ is the arc length between adjacent dipoles.  Likewise the electric dipole moments, $\mathbf{p_e}$ are given by,
\begin{equation}
\label{eq:JWeight}
|\mathbf{p_e}|=|\mathbf{J_s}| l h .
\end{equation}

At microwave frequencies, these electric and magnetic dipoles can be implemented using electrically small antennas.  For electric dipoles, electrically small straight wire antennas radiate as electric dipoles and for magnetic dipoles, electrically small wire loops radiate as magnetic dipoles. Placing these straight and loop wire antennas around a cylindrical object and feeding them with the appropriate (complex) current is sufficient to physically realize the active cloak. For the electric dipole, the dipole moment of \eqref{eq:JWeight} can be translated into an electric current by dividing $|\mathbf{p_e}|$ by the height of the cylinder $h$. For a magnetic dipole, the dipole moment $|\mathbf{p_m}|$ can be translated into an electric current on the loop using the following expression \cite{Harrington},
\begin{equation}
\label{eq:loopCurrent}
I_m=\frac{|\mathbf{p_m}|}{j \omega \mu S},
\end{equation}
where $S$ is the area of the loop in question.



\begin{figure*}[!t]
\centering
\subfloat[]
{
	\includegraphics[clip=true, trim= 3cm 1cm 9cm 3cm, scale=0.2]{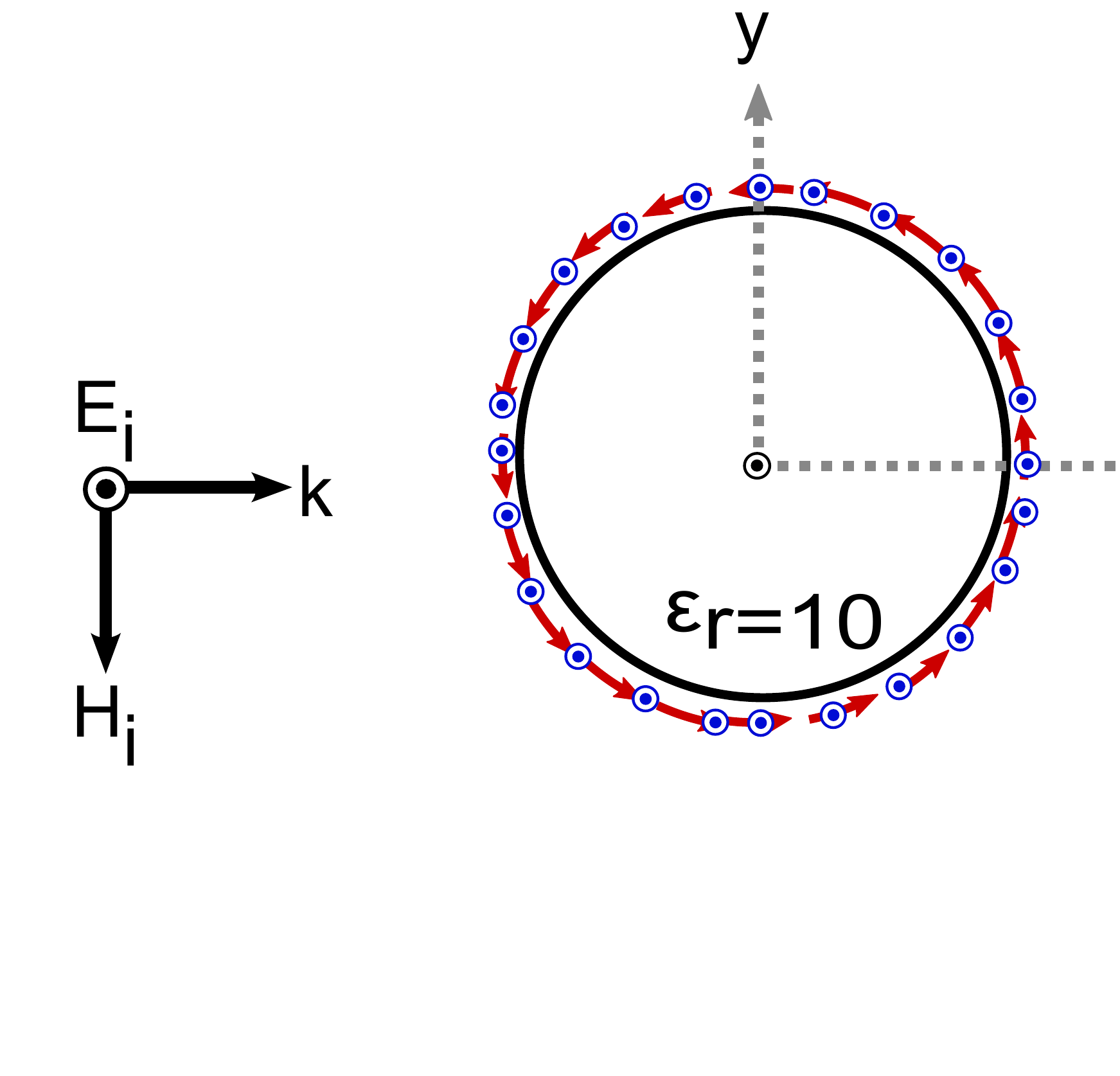}
	\label{fig:ComsolSchem}
}
\subfloat[]
{
	\includegraphics[clip=true, trim= 0cm 0cm 1cm 0cm, scale=0.2]{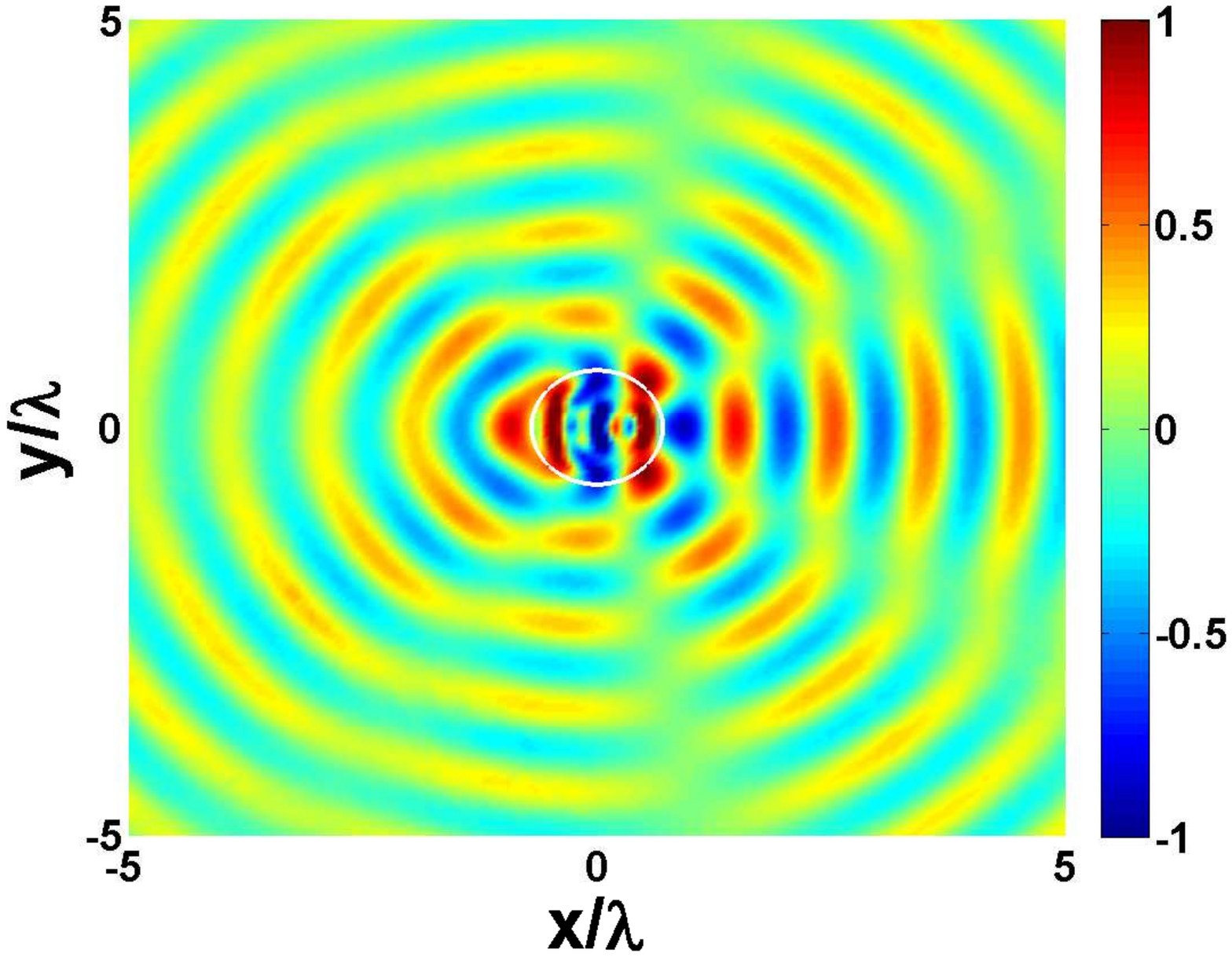}
	\label{fig:DielEscat}
}
\subfloat[]
{
	\includegraphics[clip=true, trim= 0cm 0cm 1cm 0cm, scale=0.2]{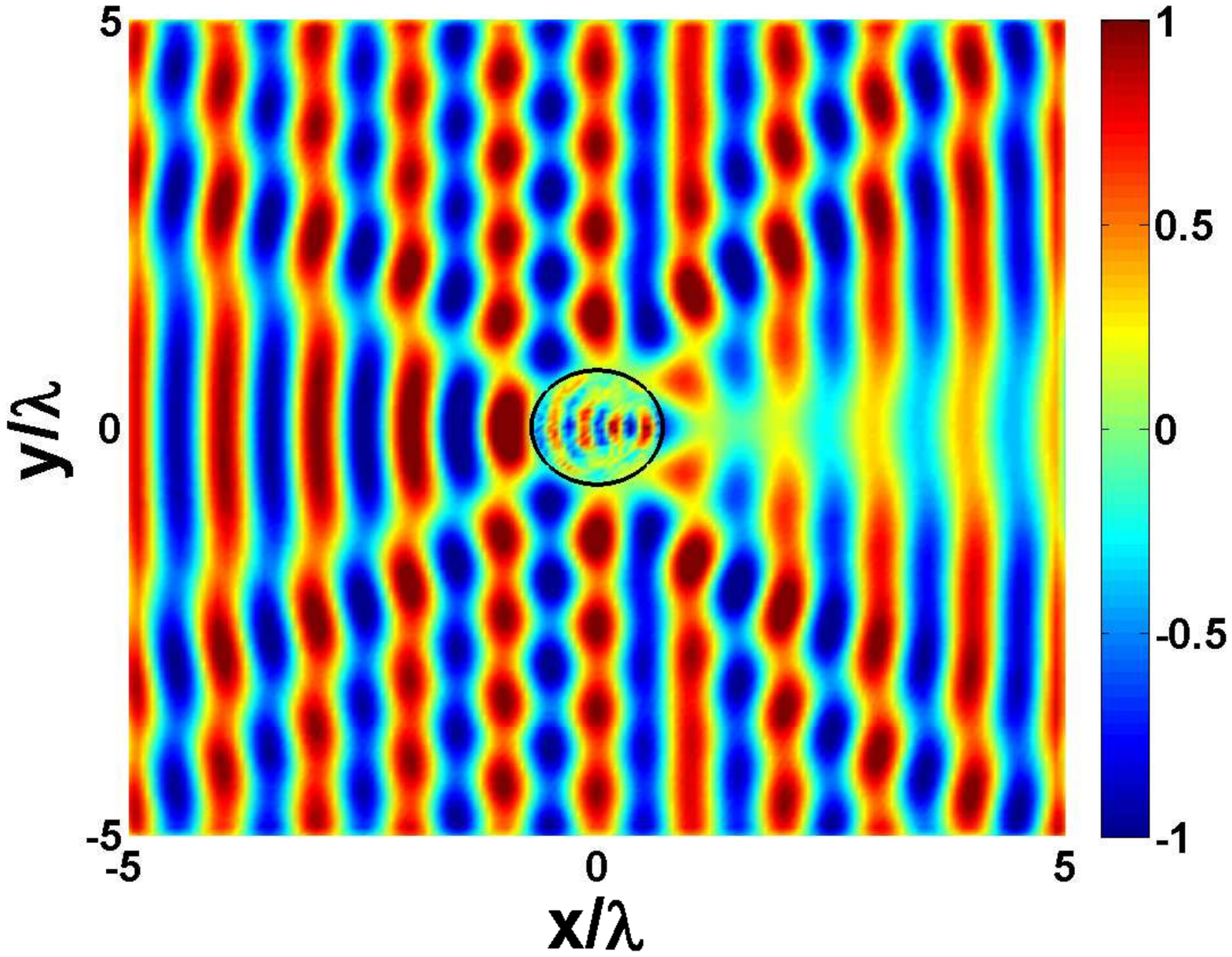}
	\label{fig:DielEtotal}
}
\\
\subfloat[]
{
	\includegraphics[clip=true, trim= 0cm 0cm 1cm 0cm, scale=0.2]{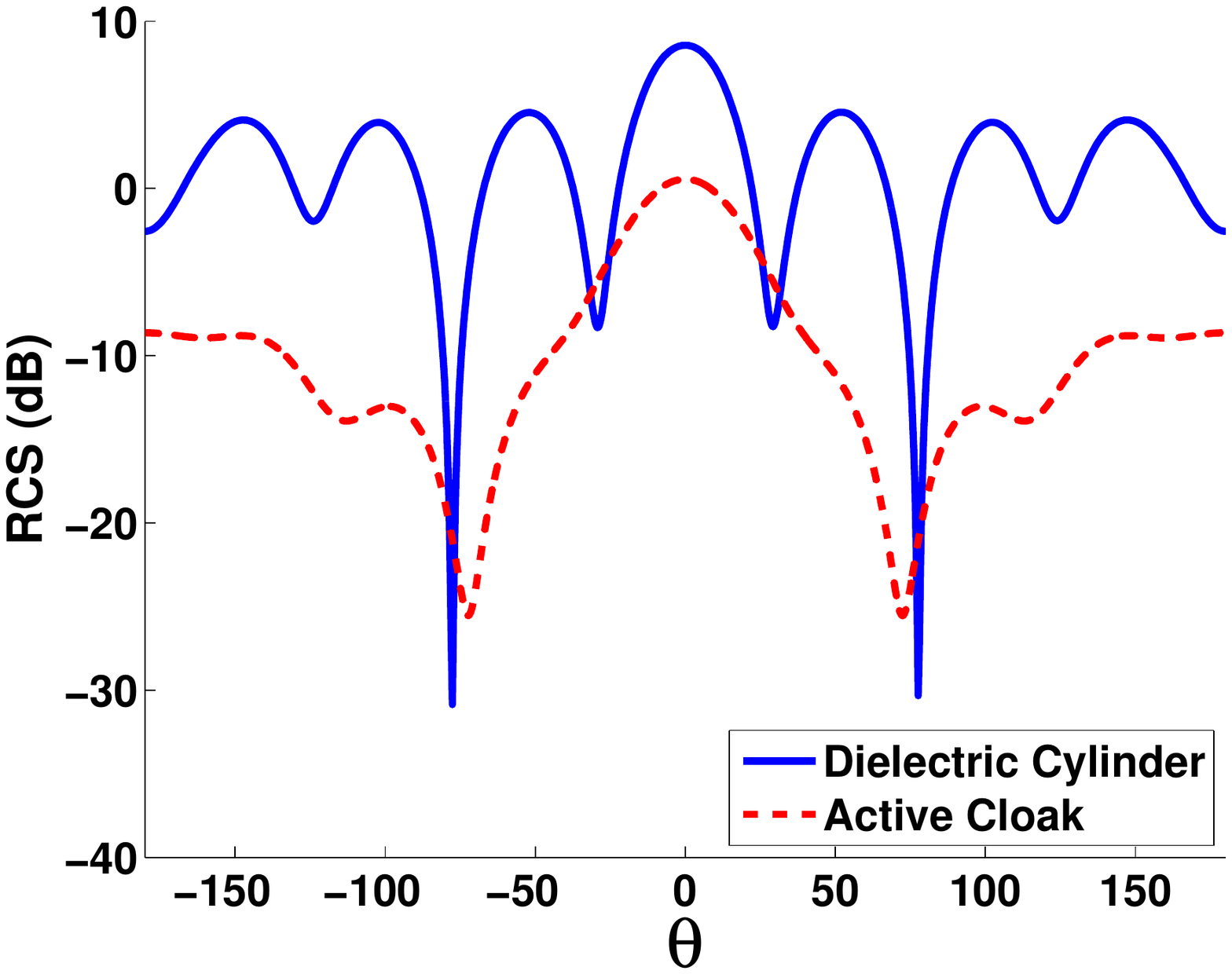}
	\label{fig:DielRCS}
}
\subfloat[]
{
	\includegraphics[clip=true, trim= 0cm 0cm 1cm 0cm, scale=0.2]{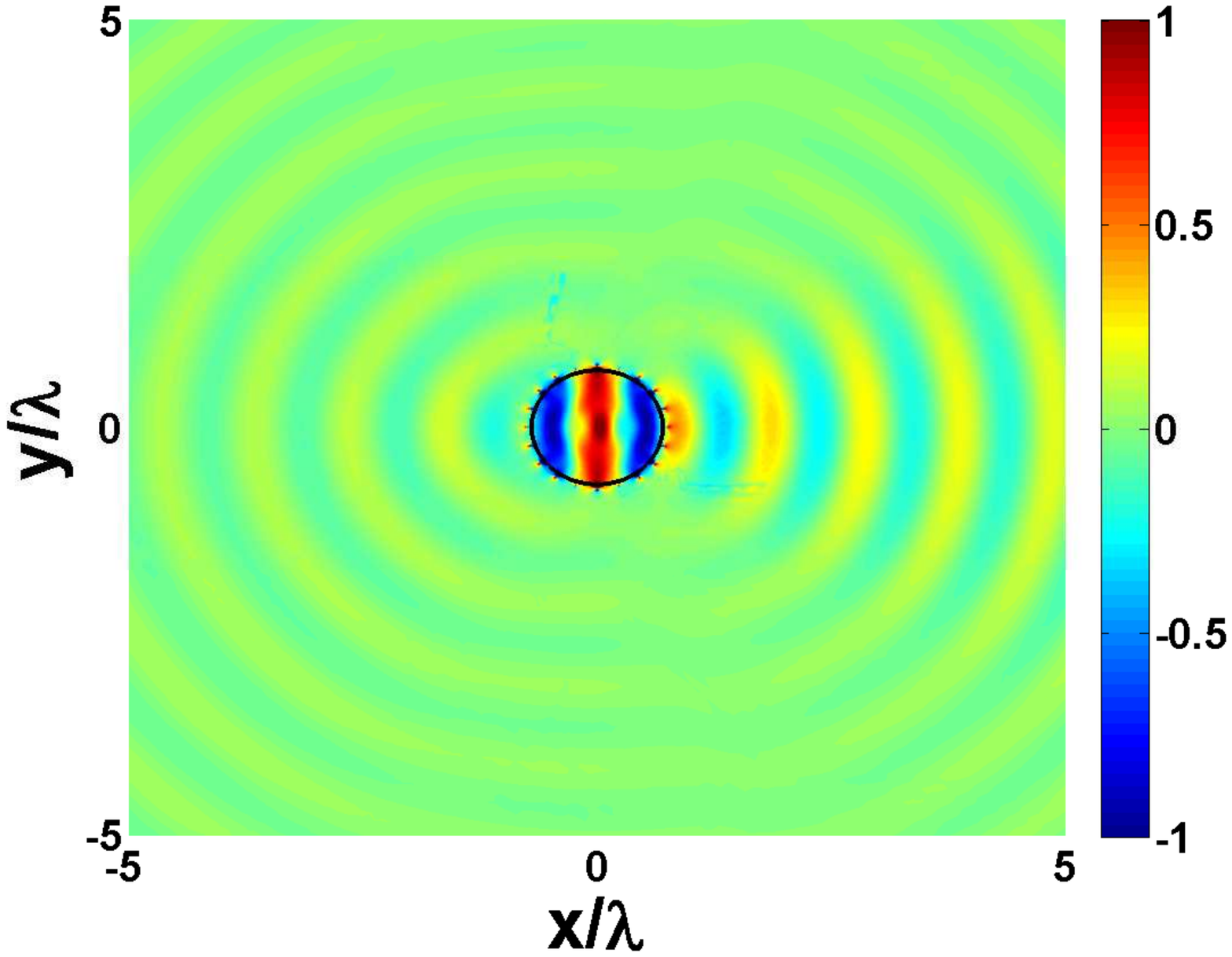}
	\label{fig:DielCloakEscat}
}
\subfloat[]
{
	\includegraphics[clip=true, trim= 0cm 0cm 1cm 0cm, scale=0.2]{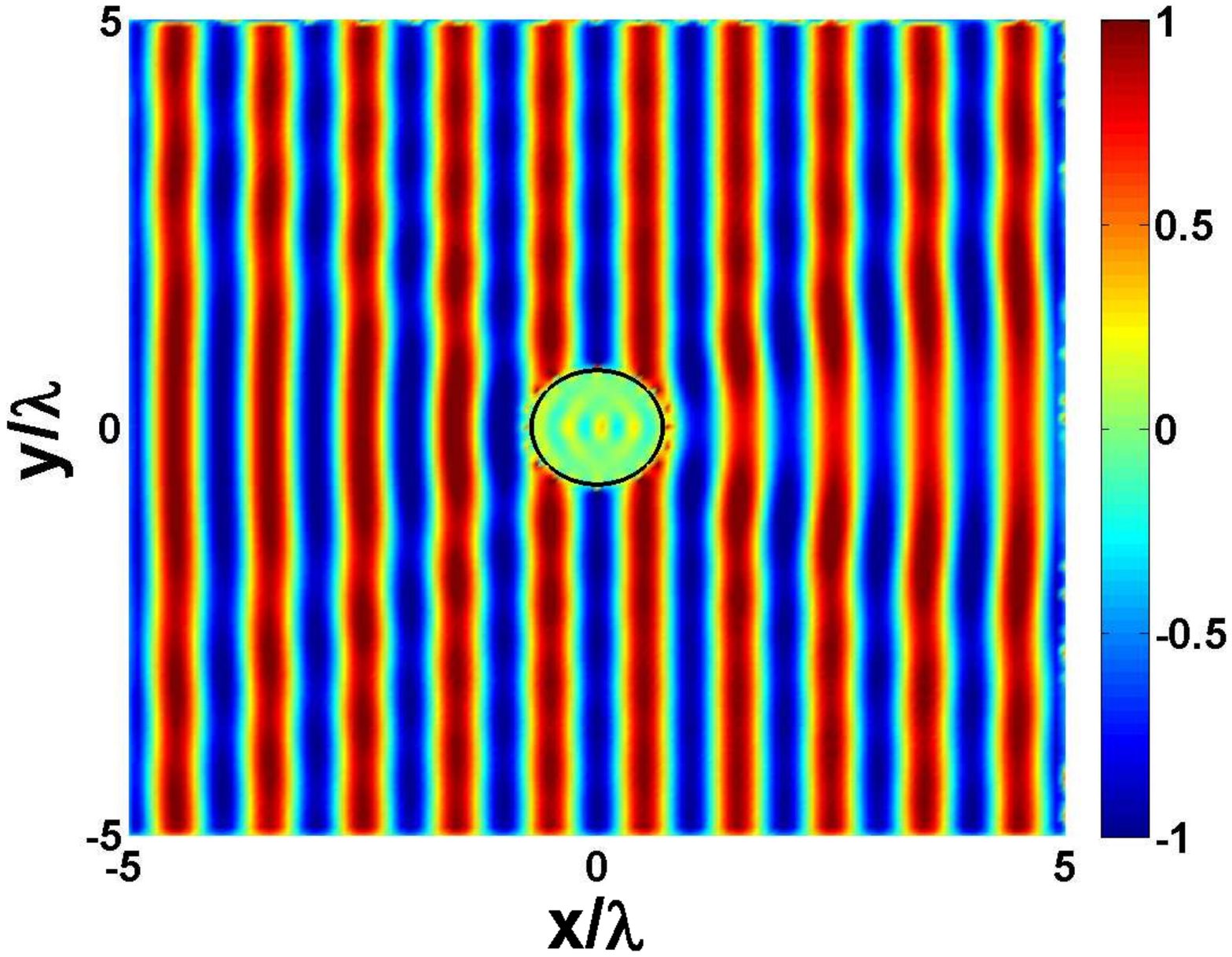}
	\label{fig:DielCloakEtotal}
}
\caption{ \protect\subref{fig:ComsolSchem} A two-dimensional dielectric cylinder surrounded by ideal electric and magnetic dipoles to cancel an incident plane wave.  \protect\subref{fig:DielEscat} The scattered electric field by the dielectric cylinder only. \protect\subref{fig:DielEtotal} The total electric field of the dielectric cylinder only. \protect\subref{fig:DielRCS} The bistatic RCS of the dielectric cylinder and the dielectric cylinder surrounded by the active cloak.  \protect\subref{fig:DielCloakEscat} The scattered electric field of the dielectric cylinder surrounded by the active cloak. \protect\subref{fig:DielCloakEtotal} The total electric field of the dielectric cylinder surrounded by the active cloak. }
\label{fig:ActiveCloakIdea}
\end{figure*}
To further demonstrate the concept of active cloaking we will use commercial electromagnetic codes to solve two examples. Both examples take place in the microwave region of the spectrum at $2.5$~GHz.  Our first example is to examine active cloaking in a purely two-dimensional environment by cloaking a circular (cylindrical) object.  We will use the commercial solver COMSOL to solve this example.  The setup of the problem  consists of a circular object with dielectric constant $\epsilon_r=10$ and radius $\rho=0.7\lambda$. A TE-polarized plane wave is incident travelling in the $\hat{x}$ direction. To characterize this cylindrical object, we have plotted the scattered electric field, the total electric field (incident electric field plus scattered electric field) and the bistatic radar cross-section (RCS) of the plane wave hitting the bare circular object in Fig.~\ref{fig:DielEscat}, Fig.~\ref{fig:DielEtotal} and Fig.~\ref{fig:DielRCS} respectively \cite{balanis}. 
From these field plots we can visualize the characteristic scattering of the dielectric cylinder and obtain a precise quantification of this scattering in terms of the RCS.

We now use COMSOL again to find the fields for the dielectric cylinder covered by an active cloak as illustrated in Fig.~\ref{fig:ComsolSchem}. Here the dielectric cylinder is surrounded by point electric and magnetic dipoles placed $\lambda/20$ away from the circular object.  For a TE-polarized plane wave the magnetic dipoles lie in the plane and the electric dipoles point out of the plane.  The weight of each electric and magnetic dipole is found by the value of the surface current density at the location of each source using \eqref{eq:MWeight} and \eqref{eq:JWeight}.  Once again the scattered electric field, the total electric field and the bistatic RCS are plotted in Fig.~\ref{fig:DielCloakEscat}, Fig.~\ref{fig:DielCloakEtotal} and Fig.~\ref{fig:DielRCS} respectively. Here we can see a noticeable change in the fields, where the scattered electric field has been reduced  and the total electric field resembles the incident plane wave. Note also that the interior fields inside the dielectric have been reduced as well.  In terms of the RCS, there is a $5.4$~dB drop in the forward scattering due to the presence of the electric and magnetic dipoles making up the active cloak.

\begin{figure*}[!t]
\centering
\subfloat[]
{
	\includegraphics[clip=true, trim= 0cm -0.5cm 0cm 0cm, scale=0.17]{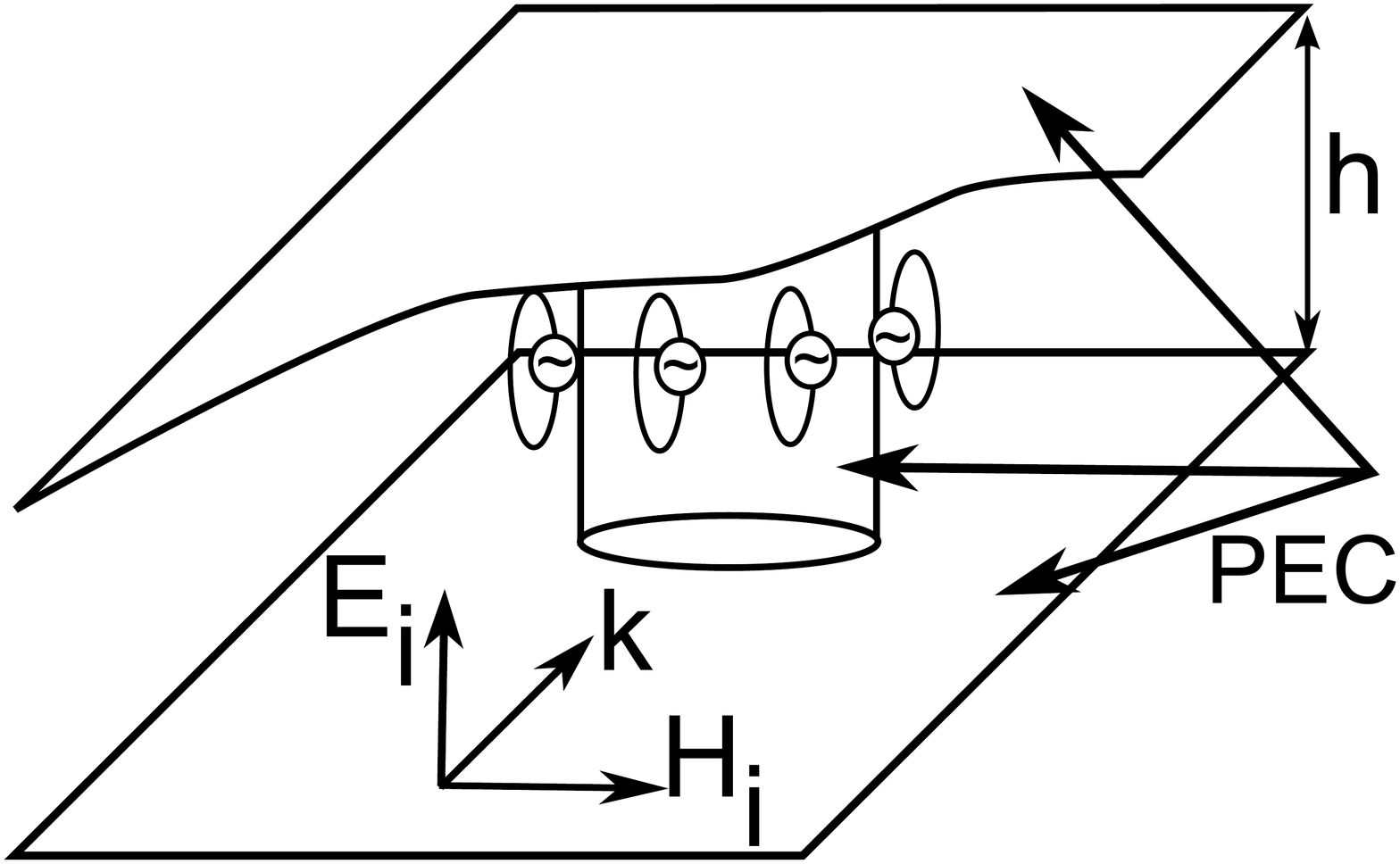}
	\label{fig:HFSSschem}
}
\subfloat[]
{
	\includegraphics[clip=true, trim= 0cm 0cm 1cm 0cm, scale=0.2]{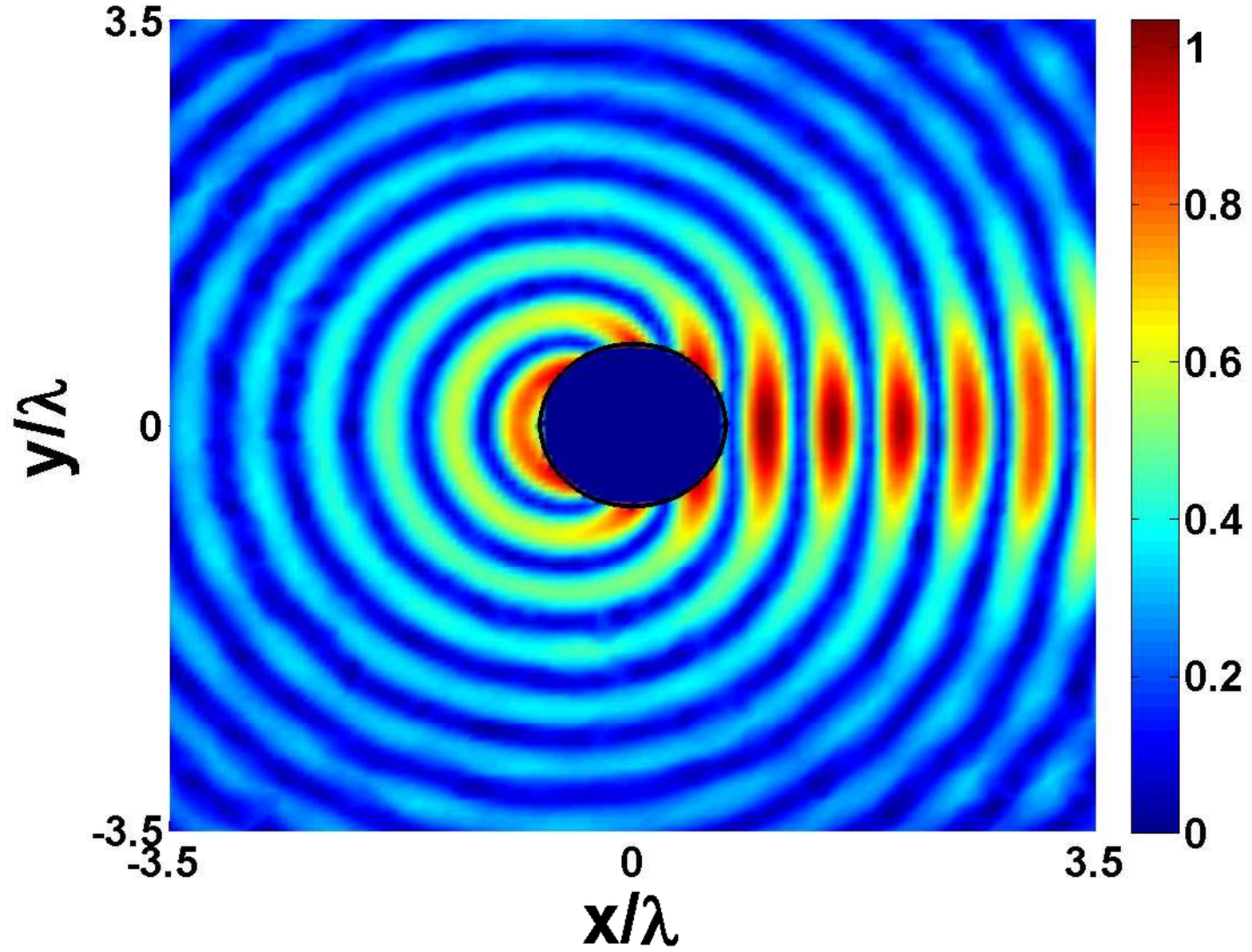}
	\label{fig:MetalEscat}
}
\subfloat[]
{
	\includegraphics[clip=true, trim= 0cm 0cm 1cm 0cm, scale=0.2]{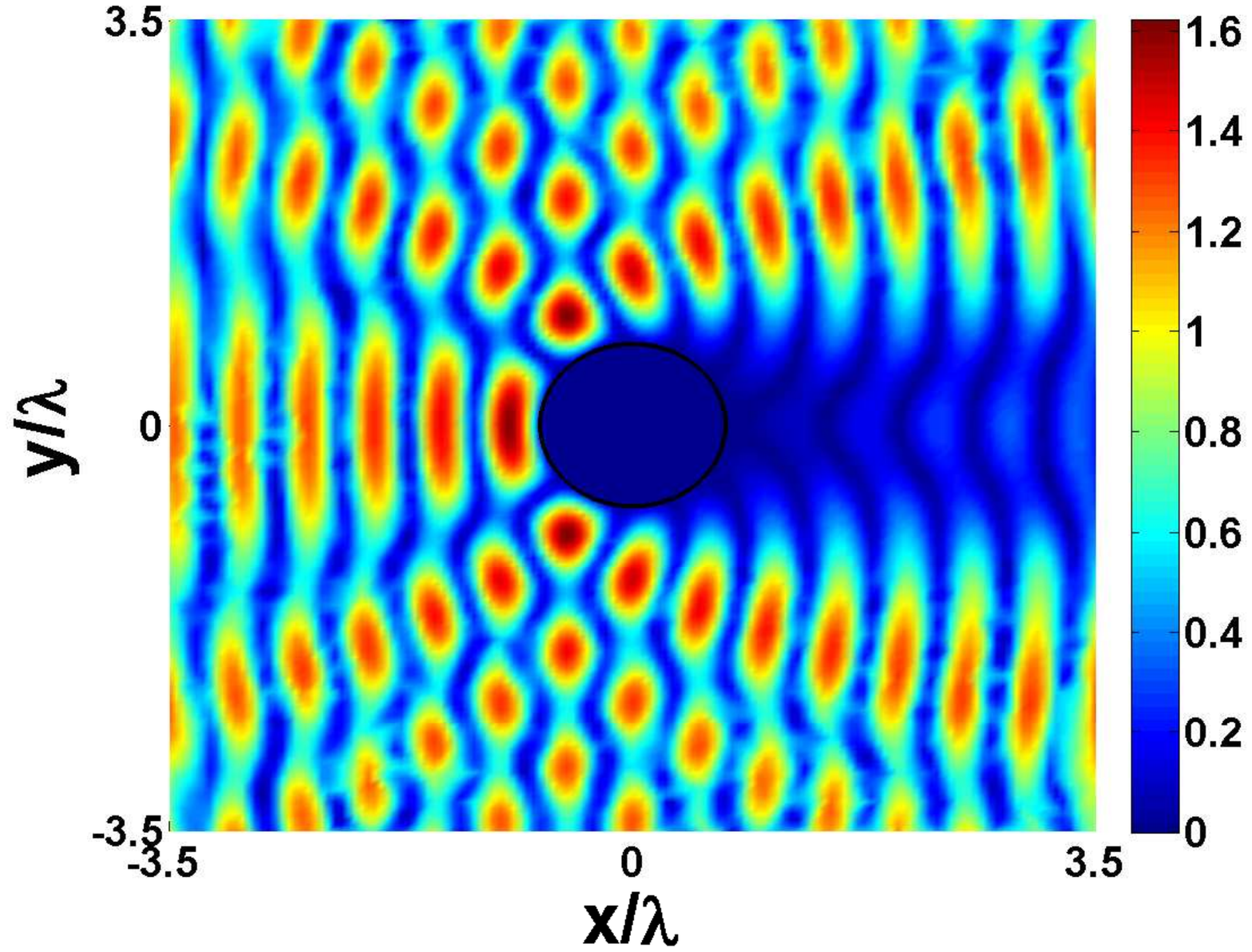}
	\label{fig:MetalEtotal}
}
\\
\subfloat[]
{
	\includegraphics[clip=true, trim= 0cm 0cm 1cm 0cm, scale=0.2]{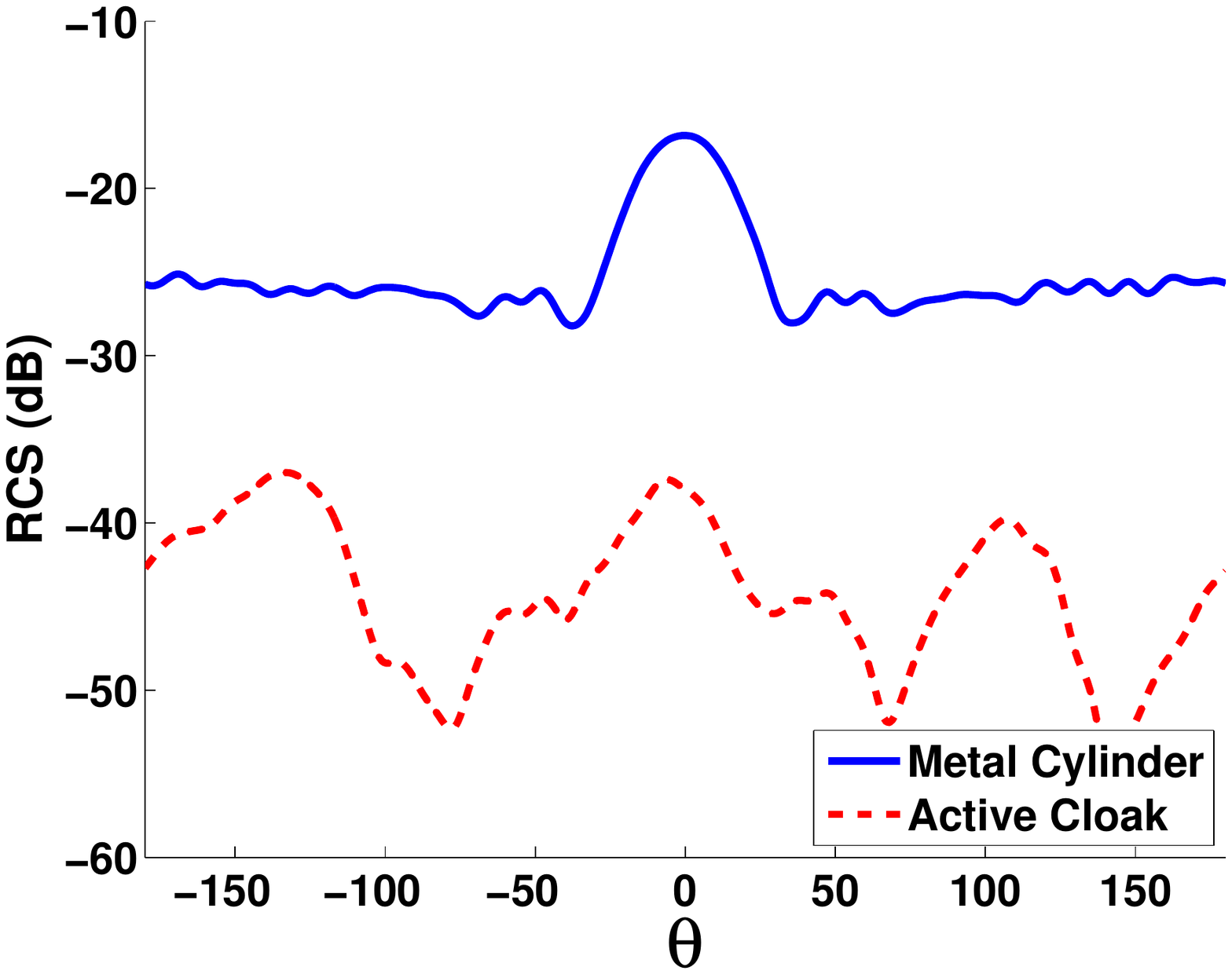}
	\label{fig:MetalRCS}
}
\subfloat[]
{
	\includegraphics[clip=true, trim= 0cm 0cm 1cm 0cm, scale=0.2]{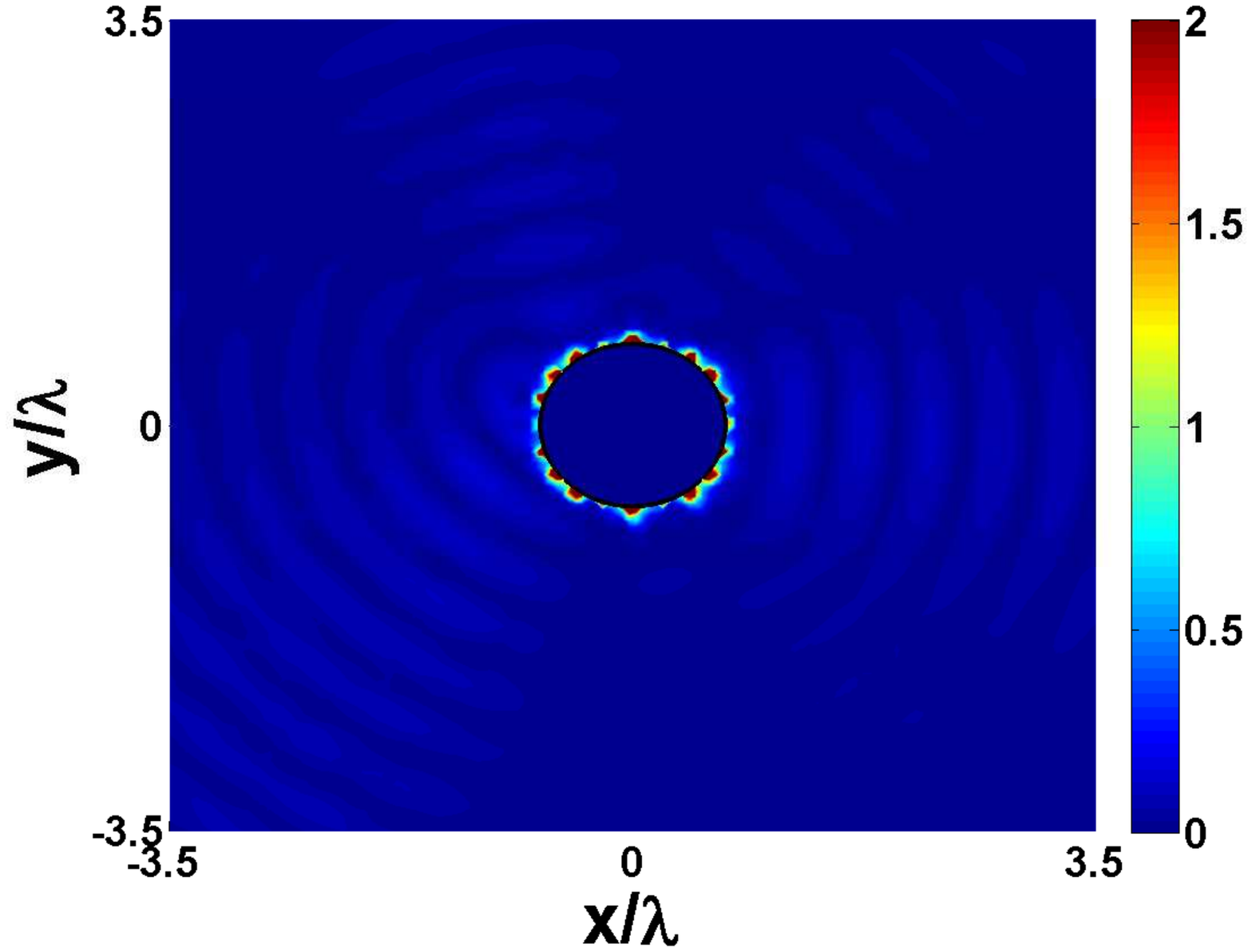}
	\label{fig:MetalCloakEscat}
}
\subfloat[]
{
	\includegraphics[clip=true, trim= 0cm 0cm 1cm 0cm, scale=0.2]{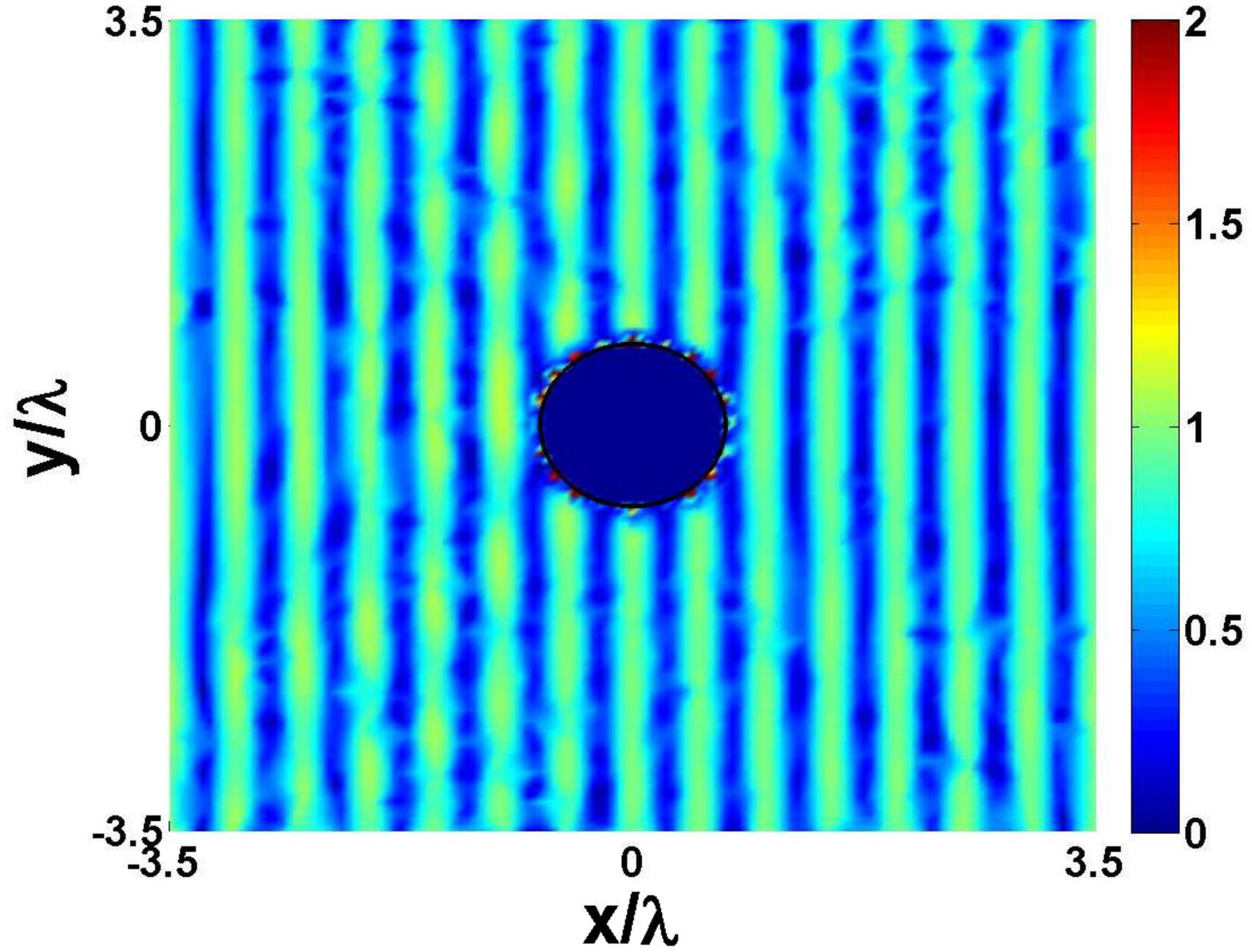}
	\label{fig:MetalCloakEtotal}
}
\caption{ \protect\subref{fig:HFSSschem} A diagram of a metal cylinder in a parallel-plate waveguide environment surrounded by an active cloak made up of loop antennas.  \protect\subref{fig:MetalEscat} The scattered electric field of the metal cylinder only. \protect\subref{fig:MetalEtotal}  The total electric field of the metal cylinder only. \protect\subref{fig:MetalRCS} The bistatic RCS of the metal cylinder and the metal cylinder surrounded by the active cloak.   \protect\subref{fig:MetalCloakEscat} The scattered electric field of the metal cylinder surrounded by the active cloak.  \protect\subref{fig:MetalCloakEtotal} The scattered electric field of the metal cylinder surrounded by the active cloak. }
\label{fig:ActiveCloakIdea}
\end{figure*}

With the feasibility of the active cloak demonstrated we now turn to a three-dimensional example that can be realized in an experimental setting.   Here we simulate, using the commercial solver Ansoft HFSS, the scattering and cloaking of a metallic cylinder placed inside a parallel-plate waveguide as shown in Fig~\ref{fig:HFSSschem}.  The purpose of the parallel-plate waveguide is to emulate an infinite domain along the $z$-axis.  This is similar to the simulated and measured environments for the transformation optics cloaks demonstrated in \cite{schurig_etal_2006}.  In this example the cylindrical object has a radius of $\rho=0.7\lambda$ but is now a perfect electrical conductor (PEC) ($\epsilon_r \rightarrow \infty$). The reason for this is that a PEC cylinder only requires magnetic dipoles to implement the active cloak as the electric dipoles are all shorted out by the presence of the PEC object.  This simplifies the implementation.  Once again the incident plane wave is TE-polarized. 

For comparison, we first examine the scattered electric field and the total electric field along  a cross-section of the waveguide as well as the bistatic RCS as plotted in Fig.~\ref{fig:MetalEscat}, Fig.~\ref{fig:MetalEtotal} and Fig.~\ref{fig:MetalRCS} respectively. Again we can see the specific signature of the forward and backward scattering of a PEC cylinder as well as a quantitative measure of it in the RCS.

To implement the active cloak, we now use  circular loops placed around the cylindrical object as illustrated in Fig.~\ref{fig:HFSSschem}. These cylindrical loops emulate a magnetic dipole and are placed at a distance of $\lambda/20$ away from the PEC object. The radius of each loop is $\lambda/20$ and each loop is fed with a current source which is weighted to implement the appropriate magnetic dipole moment as given by \eqref{eq:loopCurrent}. Simulating this environment in HFSS we can plot the scattered and total electric field as well as the RCS as shown in Fig.~\ref{fig:MetalCloakEscat}, Fig.~\ref{fig:MetalCloakEtotal} and Fig.~\ref{fig:MetalRCS} respectively. Once again, the scattered field has been reduced dramatically and the total field resembles the incident plane wave.  Likewise, the RCS shows a dramatic $20$dB drop in the forward scattering.  This demonstrates that realizable physical sources such as small antennas can be used in appropriate configurations to cancel scattered fields and can achieve good performance compared to the idealized 2D environment using point dipoles.  We note that feeding the array of sources can be accomplished using well-known antenna array designs to control the magnitude and phase of the current on each antenna \cite{balanis}.

It is worthwhile to contrast the proposed active cloak with other passive approaches to cloaking.  Compared to transformation-optics cloaking, active cloaking provides an implementation that does not rely on complex anisotropic and/or inhomogeneous media, but an array of sources which can be implemented using well known methods.  Also it can be noted that the corresponding active cloak is thin as it is only an array of wire loop antennas placed conformally around the object. Compared to plasmonic cloaking the advantage of active cloaking is the ability to cloak volumes of arbitrary size and material composition.  Finally the concept of the active cloak is, in principle, polarization independent and can be modified to accommodate multiple polarizations simply by arranging the orientation of the antennas that surround the object.

However, for active cloaking to function correctly,  the magnitude and phase of the incident field at the object must be known. This is not an insurmountable challenge as there are many practical scenarios where the fields incident on a scatterer are known, such as in communication systems where an object is obstructing an antenna \cite{Kildal_etal_1996}. In fact since our cloak is an antenna array it could potentially be used to first detect the incident field using established signal-processing algorithms and then setting the appropriate weights thus obviating this potential shortcoming \cite{Liberti_Rappaport}. 


We also make a note about the bandwidth of the active cloak.  We have demonstrated here a single-frequency solution.  To extend this solution to a band of frequencies, we note that the distribution of the incident field around the object becomes a function of frequency.  Thus at different frequencies, the required magnitude and phase of the current feeding the antennas is different.  Thus to cloak the object over a range of frequencies requires synthesizing an appropriate time-domain pulse to excite the appropriate weights on each antenna.  This can potentially make the cloak operate over large bandwidths as it circumvents the problem of routing fields around an object in a causal manner \cite{Alitalo_etal_2010}.

To summarize, we have introduced and demonstrated through numerical simulations, the concept of active cloaking.  By using an array of elementary antennas it is possible to hide an object, resulting in a thin cloak  that can be synthesized without the use of complex artificial materials and without routing the incident fields around the object to be hidden.

\bibliography{../bibtex/thesis}

\begin{thebibliography}{9}%
\makeatletter
\providecommand \@ifxundefined [1]{%
 \@ifx{#1\undefined}
}%
\providecommand \@ifnum [1]{%
 \ifnum #1\expandafter \@firstoftwo
 \else \expandafter \@secondoftwo
 \fi
}%
\providecommand \@ifx [1]{%
 \ifx #1\expandafter \@firstoftwo
 \else \expandafter \@secondoftwo
 \fi
}%
\providecommand \natexlab [1]{#1}%
\providecommand \enquote  [1]{``#1''}%
\providecommand \bibnamefont  [1]{#1}%
\providecommand \bibfnamefont [1]{#1}%
\providecommand \citenamefont [1]{#1}%
\providecommand \href@noop [0]{\@secondoftwo}%
\providecommand \href [0]{\begingroup \@sanitize@url \@href}%
\providecommand \@href[1]{\@@startlink{#1}\@@href}%
\providecommand \@@href[1]{\endgroup#1\@@endlink}%
\providecommand \@sanitize@url [0]{\catcode `\\12\catcode `\$12\catcode
  `\&12\catcode `\#12\catcode `\^12\catcode `\_12\catcode `\%12\relax}%
\providecommand \@@startlink[1]{}%
\providecommand \@@endlink[0]{}%
\providecommand \url  [0]{\begingroup\@sanitize@url \@url }%
\providecommand \@url [1]{\endgroup\@href {#1}{\urlprefix }}%
\providecommand \urlprefix  [0]{URL }%
\providecommand \Eprint [0]{\href }%
\@ifxundefined \urlstyle {%
  \providecommand \doi  [0]{\begingroup \@sanitize@url \@doi}%
  \providecommand \@doi [1]{\endgroup \@@startlink {\doibase
  #1}doi:\discretionary {}{}{}#1\@@endlink }%
}{%
  \providecommand \doi  [0]{doi:\discretionary{}{}{}\begingroup
  \urlstyle{rm}\Url }%
}%
\providecommand \doibase [0]{http://dx.doi.org/}%
\providecommand \Doi [0]{\begingroup \@sanitize@url \@Doi }%
\providecommand \@Doi  [1]{\endgroup\@@startlink{\doibase#1}\@@Doi}%
\providecommand \@@Doi [1]{#1\@@endlink}%
\providecommand \selectlanguage [0]{\@gobble}%
\providecommand \bibinfo  [0]{\@secondoftwo}%
\providecommand \bibfield  [0]{\@secondoftwo}%
\providecommand \translation [1]{[#1]}%
\providecommand \BibitemOpen [0]{}%
\providecommand \bibitemStop [0]{}%
\providecommand \bibitemNoStop [0]{.\EOS\space}%
\providecommand \EOS [0]{\spacefactor3000\relax}%
\providecommand \BibitemShut  [1]{\csname bibitem#1\endcsname}%
\bibitem [{\citenamefont {Schurig}\ \emph {et~al.}(2006)\citenamefont
  {Schurig}, \citenamefont {Mock}, \citenamefont {Justice}, \citenamefont
  {Cummer}, \citenamefont {Pendry}, \citenamefont {Starr},\ and\ \citenamefont
  {Smith}}]{schurig_etal_2006}%
  \BibitemOpen
  \bibfield  {author} {\bibinfo {author} {\bibfnamefont {D.}~\bibnamefont
  {Schurig}}, \bibinfo {author} {\bibfnamefont {J.~J.}\ \bibnamefont {Mock}},
  \bibinfo {author} {\bibfnamefont {B.~J.}\ \bibnamefont {Justice}}, \bibinfo
  {author} {\bibfnamefont {S.~A.}\ \bibnamefont {Cummer}}, \bibinfo {author}
  {\bibfnamefont {J.~B.}\ \bibnamefont {Pendry}}, \bibinfo {author}
  {\bibfnamefont {A.~F.}\ \bibnamefont {Starr}}, \ and\ \bibinfo {author}
  {\bibfnamefont {D.~R.}\ \bibnamefont {Smith}},\ }\Doi
  {10.1126/science.1133628} {\bibfield  {journal} {\bibinfo  {journal}
  {Science},\ }\textbf {\bibinfo {volume} {314}},\ \bibinfo {pages} {977}
  (\bibinfo {year} {2006})}\BibitemShut {NoStop}%
\bibitem [{\citenamefont {Al\'u}\ and\ \citenamefont
  {Engheta}(2008)}]{Alu_Engheta_2008}%
  \BibitemOpen
  \bibfield  {author} {\bibinfo {author} {\bibfnamefont {A.}~\bibnamefont
  {Al\'u}}\ and\ \bibinfo {author} {\bibfnamefont {N.}~\bibnamefont
  {Engheta}},\ }\href@noop {} {\bibfield  {journal} {\bibinfo  {journal}
  {Journal of Optics A: Pure and Applied Optics},\ }\textbf {\bibinfo {volume}
  {10}},\ \bibinfo {pages} {093002} (\bibinfo {year} {2008})}\BibitemShut
  {NoStop}%
\bibitem [{\citenamefont {Alitalo}\ and\ \citenamefont
  {Tretyakov}(2011)}]{Alitalo_Tretyakov_2011}%
  \BibitemOpen
  \bibfield  {author} {\bibinfo {author} {\bibfnamefont {P.}~\bibnamefont
  {Alitalo}}\ and\ \bibinfo {author} {\bibfnamefont {S.}~\bibnamefont
  {Tretyakov}},\ }\Doi {10.1109/JPROC.2010.2093471} {\bibfield  {journal}
  {\bibinfo  {journal} {Proceedings of the IEEE},\ }\textbf {\bibinfo {volume}
  {99}},\ \bibinfo {pages} {1646 } (\bibinfo {year} {2011})},\ ISSN \bibinfo
  {issn} {0018-9219}\BibitemShut {NoStop}%
\bibitem [{\citenamefont {Kildal}\ \emph {et~al.}(1996)\citenamefont {Kildal},
  \citenamefont {Kishk},\ and\ \citenamefont {Tengs}}]{Kildal_etal_1996}%
  \BibitemOpen
  \bibfield  {author} {\bibinfo {author} {\bibfnamefont {P.-S.}\ \bibnamefont
  {Kildal}}, \bibinfo {author} {\bibfnamefont {A.}~\bibnamefont {Kishk}}, \
  and\ \bibinfo {author} {\bibfnamefont {A.}~\bibnamefont {Tengs}},\ }\Doi
  {10.1109/8.542076} {\bibfield  {journal} {\bibinfo  {journal} {Antennas and
  Propagation, IEEE Transactions on},\ }\textbf {\bibinfo {volume} {44}},\
  \bibinfo {pages} {1509 } (\bibinfo {year} {1996})},\ ISSN \bibinfo {issn}
  {0018-926X}\BibitemShut {NoStop}%
\bibitem [{\citenamefont {Harrington}(2001)}]{Harrington}%
  \BibitemOpen
  \bibfield  {author} {\bibinfo {author} {\bibfnamefont {R.~F.}\ \bibnamefont
  {Harrington}},\ }\href@noop {} {\emph {\bibinfo {title} {Time-Harmonic
  Electromagnetic Fields}}}\ (\bibinfo  {publisher} {Wiley-Interscience},\
  \bibinfo {year} {2001})\BibitemShut {NoStop}%
\bibitem [{\citenamefont {Oppenheim}\ \emph {et~al.}(1997)\citenamefont
  {Oppenheim}, \citenamefont {Willsky},\ and\ \citenamefont
  {Nawab}}]{Oppenheim}%
  \BibitemOpen
  \bibfield  {author} {\bibinfo {author} {\bibfnamefont {A.~V.}\ \bibnamefont
  {Oppenheim}}, \bibinfo {author} {\bibfnamefont {A.~S.}\ \bibnamefont
  {Willsky}}, \ and\ \bibinfo {author} {\bibfnamefont {S.~H.}\ \bibnamefont
  {Nawab}},\ }\href@noop {} {\emph {\bibinfo {title} {Signals and Systems}}}\
  (\bibinfo  {publisher} {Prentice Hall},\ \bibinfo {year} {1997})\BibitemShut
  {NoStop}%
\bibitem [{\citenamefont {Balanis}(2005)}]{balanis}%
  \BibitemOpen
  \bibfield  {author} {\bibinfo {author} {\bibfnamefont {C.~A.}\ \bibnamefont
  {Balanis}},\ }\href@noop {} {\emph {\bibinfo {title} {Antenna Theory:
  Analysis and Design}}}\ (\bibinfo  {publisher} {Wiley-Interscience},\
  \bibinfo {year} {2005})\ ISBN \bibinfo {isbn} {0471714623}\BibitemShut
  {NoStop}%
\bibitem [{\citenamefont {Liberti}\ and\ \citenamefont
  {Rappaport}(1999)}]{Liberti_Rappaport}%
  \BibitemOpen
  \bibfield  {author} {\bibinfo {author} {\bibfnamefont {J.}~\bibnamefont
  {Liberti}}\ and\ \bibinfo {author} {\bibfnamefont {T.~S.}\ \bibnamefont
  {Rappaport}},\ }\href@noop {} {\emph {\bibinfo {title} {Smart Antennas for
  Wireless Communications: IS-95 and Third Generation CDMA Applications}}}\
  (\bibinfo  {publisher} {Prentice Hall},\ \bibinfo {year} {1999})\BibitemShut
  {NoStop}%
\bibitem [{\citenamefont {Alitalo}\ \emph {et~al.}(2010)\citenamefont
  {Alitalo}, \citenamefont {Kettunen},\ and\ \citenamefont
  {Tretyakov}}]{Alitalo_etal_2010}%
  \BibitemOpen
  \bibfield  {author} {\bibinfo {author} {\bibfnamefont {P.}~\bibnamefont
  {Alitalo}}, \bibinfo {author} {\bibfnamefont {H.}~\bibnamefont {Kettunen}}, \
  and\ \bibinfo {author} {\bibfnamefont {S.}~\bibnamefont {Tretyakov}},\
  }\href@noop {} {\bibfield  {journal} {\bibinfo  {journal} {Journal of Applied
  Physics},\ }\textbf {\bibinfo {volume} {107}},\ \bibinfo {eid} {034905}
  (\bibinfo {year} {2010})}\BibitemShut {NoStop}%
\end{thebibliography}%

\end{document}